\documentclass[prb,aps,onecolumn,showpacs,floats,a4paper,times,12pt]{revtex4}
\usepackage{epsfig,bm,epsf,graphics}
\begin{document}
\draft
\title{Stacked Triangular XY Antiferromagnets: End of a Controversial Issue on the Phase Transition}
\author{V. Thanh Ngo$^{a,b}$ and H. T. Diep\footnote{ Corresponding author, E-mail:diep@u-cergy.fr }}
\address{Laboratoire de Physique Th\'eorique et Mod\'elisation,
CNRS-Universit\'e de Cergy-Pontoise, UMR 8089\\
2, Avenue Adolphe Chauvin, 95302 Cergy-Pontoise Cedex, France\\
$^a$ Institute of Physics, P.O. Box 429,   Bo Ho, Hanoi 10000,
Vietnam\\
$^b$ Asia Pacific Center for Theoretical Physics, Hogil Kim
Memorial Building 5th floor, POSTECH, Hyoja-dong, Namgu, Pohang
790-784, Korea}

\begin{abstract}
We show in this paper by using the Wang-Landau flat-histogram Monte
Carlo method that the phase transition in the XY stacked triangular
antiferromagnet is clearly of first-order, confirming results from
latest Monte Carlo simulation and from a nonperturbative
renormalization group, putting an end to a long-standing
controversial issue.
\end{abstract}
\maketitle

\section{Introduction}

Effects of the frustration in spin systems have been extensively
investigated during the last 30 years.  Frustrated spin systems are shown to
have unusual properties such as large ground state (GS)
degeneracy, additional GS symmetries, successive phase transitions
with complicated nature.  Frustrated systems still challenge
theoretical and experimental methods. For  recent reviews, the
reader is referred to Ref. \onlinecite{Diep2005}.

Let us confine our discussion on the nature of the phase transition
in strongly frustrated spin systems.  Since the nature of the phase
transition depends on the symmetry and the dimension of the system,
we have to examine first its GS properties.  Of course, the nature
of the order parameter defined according to the system symmetry
determines the properties of the phase transition. One of the most
studied systems is the stacked triangular antiferromagnet (STA): the
antiferromagnetic (AF) interaction between nearest-neighbor (NN)
spins on the triangular lattice causes  a very strong frustration.
It is impossible\cite{Diep2005} to fully satisfy simultaneously the
three AF bond interactions on each equilateral triangle.  The GS
configuration of both Heisenberg and XY models is the well-known
120-degree structure.

The phase transition in strongly frustrated spin systems is a
subject of intensive investigations in the last 20 years.
Theoretically, these systems are excellent testing grounds for
theories and approximations.  Many well-established methods such as
renormalization group (RG), high- and low-temperature series
expansions etc often failed to deal with these systems.
Experimentally, data on different frustrated systems show a variety
of possibilities: first-order or second-order transitions with
unknown critical exponents etc. (see reviews in Ref.
\onlinecite{Diep2005}). The case of XY and Heisenberg spins on the
STA has been intensively studied mostly since 1987.  There are good
recent reviews on the subject (see for example reviews by Delamotte
et al\cite{Delamotte2004}). Let us briefly recall here some main
historical developments and actual situation. In the XY and
Heisenberg cases, different materials give different experimental
results. The anomalous dimension is found negative in many materials
and in most numerical simulations, the scaling relations are
violated and no universality in the exponents was found in early
simulations.  This situation is briefly described in the following.
Kawamura~\cite{kawamura87,kawamura88} has  conjectured by the
two-loop RG analysis  in $d=3$  the existence of a new universality
class for frustrated magnets. Since then there have been many other
perturbative calculations with contradictory
results.\cite{azaria90,antonenko95} From 2000, there has been a
number of papers  by Tessier and coworkers
~\cite{tissier00b,tissier00,tissier01} using a nonperturbative RG
study of frustrated magnets for any dimension between  two and four.
They recovered all known perturbative results at one loop in two and
four dimensions as well as for $N\to \infty$. They determined
$N_c(d)$ for all $d$ and found $N_c(d=3)=5.1$ below which the
transition is of first order. However, they found the existence of a
whole region in the flow diagram in which the flow is slow. As a
consequence, for $N=2,3$, they found pseudo-critical exponents in
good agreement with some experimental data. This allowed them to
account for the nonuniversal scaling observed in XY and Heisenberg
frustrated magnets.  The  only problem in this nonperturbative
technique is that the Hamiltonian is truncated at the beginning.
Giving this fact, we have to be careful about its conclusion. As
will be seen in this paper, the nonperturbative results are so far
well confirmed. Early MC results on XY STA have been reviewed by
Loison.\cite{Loison}  Until 2003, all numerical simulations found
ambiguous results for this model and never a clear first-order
transition.  A numerical breakthrough has been realized with the
results of Itakura\cite{itakura03} who used an improved MC
renormalization-group scheme to numerically investigate the
renormalization group flow of the
  Heisenberg and XY STA and its
  effective Hamiltonian which is used in the
  field-theoretical studies. He found that the XY STA
  exhibits clear first-order behavior and there are no chiral
  fixed points of renormalization-group
  flow for $N$=2 and 3 cases.
In 2004, Peles et al\cite{Peles} have used a continuous model to
study the XY STA by MC simulation. They found evidence of a
first-order transition.  In 2006, Kanki et al\cite{Kanki}, using a
microcanonical MC method, have found a first-order signature of the
XY STA. While these recent simulations have demontrated evidence of
first-order transition for the XY STA in agreement with the
nonperturbative RG analysis, all of them suffer one or two uncertain
aspects: the work of Itakura has used a truncated Hamiltonian, the
work of Peles et al has used standard MC methods and the work of
Kanki et al used a traditional microcanonical MC technique. At
present, we have a very high-performance technique at hand for weak
first-order transitions.  This is a very good opportunity to test it
on the XY STA and to say a last word on the nature of the phase
transition of this system by using the full Hamiltonian, confirming
or rejecting the nonperturbative RG and recent MC results. That is
the purpose of this work.

We study again here the XY STA with high-resolution MC technique
which is very efficient specially for weak first-order
transition.\cite{WL1} Our aim is to try to put an end to the
controversy which has been lasting for 20 years. We will recall some
important numerical results in the next section.

The paper is organized as follows. Section II is devoted to the
description of the model and  technical details of the Wang-Landau
(WL) methods as applied in the present paper.  Section III shows our
results.  Concluding remarks are given in section IV.

\section{Monte Carlo Simulation: Wang-Landau algorithm}

We consider the stacking of triangular lattices in the $z$
direction.  The spins are the classical XY model of magnitude
$S=1$.  The Hamiltonian is given by

\begin{equation}
\mathcal H=J\sum_{\left<i,j\right>}\mathbf S_i\cdot\mathbf S_j
+J'\sum_{\left<i,k\right>}\mathbf S_i\cdot\mathbf S_k
\label{eqn:hamil1}
\end{equation}
where $\mathbf S_i$ is the XY spin at the lattice site $i$,
$\sum_{\left<i,j\right>}$ indicates the sum over the NN spin pairs
$\mathbf S_i$ and $\mathbf S_j$ in a $xy$ triangular plane, while
$\sum_{\left<i,k\right>}$ indicates that of NN spin pairs between
adjacent planes.  $J$ and $J'$ are in-plane and inter-plane
interactions, respectively. We shall suppose that $J=1$
(antiferromagnetic) and $J'=-1$ (ferromagnetic) in the following.

%%%%%%%%%%%

Recently, Wang and Landau\cite{WL1} proposed a Monte Carlo
algorithm for classical statistical models. The algorithm uses a
random walk in energy space in order to obtained an accurate
estimate for the density of states $g(E)$. This method is based on
the fact that a flat energy histogram $H(E)$ is produced if the
probability for the transition to a state of energy $E$ is
proportional to $g(E)^{-1}$.

At the beginning of the simulation, the density of states (DOS) is
set equal to one for all energies, $g(E)=1$. In general, if $E$
and $E'$ are the energies before and after a spin is flipped, the
transition probability from $E$ to $E'$ is
\begin{equation}
p(E\rightarrow E')=\min\left[g(E)/g(E'),1\right].
\label{eq:wlprob}
\end{equation}

Each time an energy level $E$ is visited, the DOS is modified by a
modification factor $f>0$ whether the spin flipped or not, i.e.
$g(E)\rightarrow g(E)f$.
 In the beginning of the random walk the modification factor $f$ can be
as large as $e^1\simeq 2.7182818$. A histogram $H(E)$ records how
often a state of energy $E$ is visited. Each time the energy
histogram satisfies a certain "flatness" criterion, $f$ is reduced
according to $f\rightarrow \sqrt{f}$ and $H(E)$ is reset to zero
for all energies. The reduction process of the modification factor
$f$ is repeated several times until a final value
$f_{\mathrm{final}}$ which close enough to one. The histogram is
considered as flat if
\begin{equation}
H(E)\ge x\%.\langle H(E)\rangle \label{eq:wlflat}
\end{equation}
for all energies, where the flatness parameter $0\% < x\%<100\%$
controls the accuracy of the estimated $g(E)$, with increasing
accuracy as $x\%$ approaches unity. $\langle H(E)\rangle $ is the
average histogram.

Thermodynamic quantities\cite{WL1,brown} can be evaluated using the
canonical distribution at any temperature $T$ by
%\begin{equation}
$P(E,T) =g(E)\exp(-E/k_BT)/Z$ where $Z$ is the partition function
defined by $Z =\sum_E g(E)\exp(-E/k_BT)$.

%\begin{eqnarray}
%\langle E^n\rangle &=&\frac{1}{Z}\sum_E E^n g(E)\exp(-E/k_BT),\\
%C_v&=&\frac{\langle E^2\rangle-\langle E\rangle^2}{k_BT^2},\\
%\langle M^n\rangle &=&\frac{1}{Z}\sum_E M^n g(E)\exp(-E/k_BT),\\
%\chi&=&\frac{\langle M^2\rangle-\langle M\rangle^2}{k_BT},
%\end{eqnarray}
%where $Z$ is the partition function and defined by
%\begin{equation}
%Z =\sum_E g(E)\exp(-E/k_BT). \label{eq:partfunc}
%\end{equation}

In this work, we consider a energy range of
interest\cite{Schulz,Malakis} $(E_{\min},E_{\max})$. We divide this
energy range to $R$ subintervals, the minimum energy of each
subinterval is $E^i_{\min}$ for $i=1,2,...,R$, and maximum of the
subinterval $i$ is $E^i_{\max}=E^{i+1}_{\min}+2\Delta E$, where
$\Delta E$ can be chosen large enough for a smooth boundary between
two subintervals. The Wang-Landau algorithm is used to calculate the
relative DOS of each subinterval $(E^i_{\min},E^i_{\max})$ with the
modification factor $f_\mathrm{final}=\exp(10^{-9})$ and flatness
criterion $x\%=95\%$. We reject the suggested spin flip and do not
update $g(E)$ and the energy histogram $H(E)$ of the current energy
level $E$ if the spin-flip trial would result in an energy outside
the energy segment. The DOS of the whole range is obtained by
joining the DOS of each subinterval $(E^i_{\min}+\Delta
E,E^i_{\max}-\Delta E)$.

\section{Results}

 We used the system size of $N\times N \times N$ where
$N=12,18,24, 30,36, 48, 60, 72,84, 90,96, 108,120$. Periodic
boundary conditions are used in the three directions.  $|J|=1$ is
taken as unit of energy in the following.

\begin{figure}
\centerline{\epsfig{file=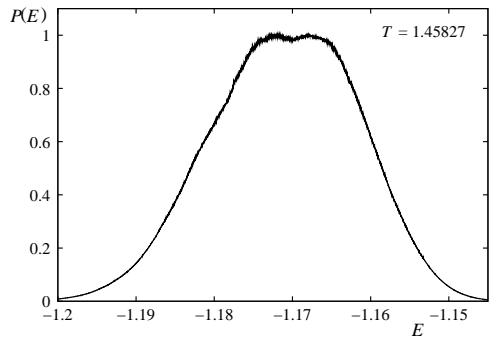,width=2.5in}} \caption{Energy
histograms for $N=48$ at $T_c$ indicated on the
figure.}\label{fig:STA48PE}
\end{figure}

\begin{figure}
\centerline{\epsfig{file=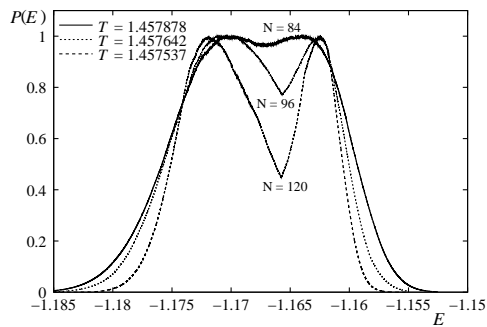,width=2.5in}} \caption{Energy
histograms for  $N=84,96,120$ at $T_c$ indicated on the figure.}
\label{fig:STA120PE}
\end{figure}

\begin{figure}
\centerline{\epsfig{file=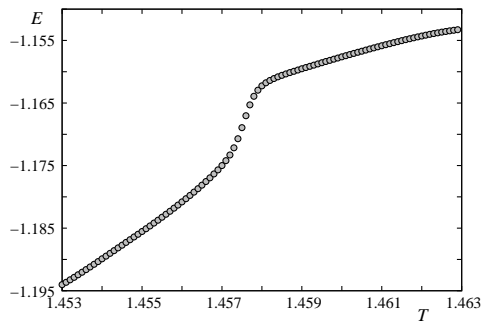,width=2.5in}} \caption{Energy
 vs $T$ for $N=120$.} \label{fig:STA120E}
\end{figure}

The energy histograms for two representative sizes $N=48$ and
$N=120$ are shown in Figs. \ref{fig:STA48PE} and \ref{fig:STA120PE},
respectively.  As seen, for $N=48$, the peak, though very large,
does not show yet a double-maximum structure.  Only from $N=90$ that
the double-peak structure clearly appears.  This is a sufficient
condition, not a necessary condition, for a first-order transition.
We give here the values of $T_c$ for a few sizes:  $T_c =
1.458270,1.457878, 1.457642, 1.457537$ for $N = 48,84,96,120$,
respectively.  Note that this result is in excellent agreement with
earlier MC simulations\cite{itakura03,Peles,Kanki} using less
sophisticated methods.  To explain why standard MC methods  without
histogram monitoring (see for example Ref. \onlinecite{kawamura87})
fail to see the first order character, let us show in Fig.
\ref{fig:STA120E} the energy vs $T$ obtained by averaging over
states obtained by the WL method for $N=120$. We see here that even
at this big size, the average energy does not show a discontinuity
as in a strong first-order transition: the averaging over all states
erases away the bimodal distribution seen  in the energy histogram
at the transition temperature.  Therefore care should be taken to
avoid such problems due to averaging in MC simulations.  We note
that the distance between to peaks in Fig. \ref{fig:STA120PE}, i. e.
the latent heat, is $\simeq 0.009$ in agreement with earlier
works.\cite{itakura03,Peles,Kanki}

%\begin{figure}
%\centerline{\epsfig{file=STA36PE.eps,width=2.8in}} \caption{Energy
%histogram for $N=36$.} \label{fig:STA36PE}
%\end{figure}

%\begin{figure}
%\centerline{\epsfig{file=STA36GE.eps,width=2.8in}}
%\caption{Density of state for $N=36$.} \label{fig:STA36GE}
%\end{figure}


\section{Concluding Remarks}

We have studied in this paper the phase transition in the XY STA by
using the flat histogram technique invented by Wang and Landau. The
method is very efficient because it helps to overcome extremely long
transition time between energy valleys in systems with a possible
first-order phase transition.   We found that the transition is
clearly of first-order confirming therefore recent MC results using
less efficient techniques.  These results put definitely an end to
the 20-year long controversy and lend support to nonperturbative RG
calculations using an effective average Hamiltonian.

{}

\end{document}